# Electron Attachment to DNA: The Protective Role of Amino Acids

Pooja Verma and Achintya Kumar Dutta[*]

*Department of Chemistry, Indian Institute of Technology Bombay, Powai, Mumbai-400076*

**Abstract:**

We have studied the effect of amino acids on electron attachment properties of DNA nucleobases, taking cytosine as a model system. The equation of motion coupled cluster theory with an extended basis set has been used to simulate the electron-attached state of the DNA model system. Four selected amino acids, Arginine, Alanine, Lysine, and Glycine which form a major component of histone proteins are considered to investigate their role in electron attachment to DNA nucleobase. The electron attachment to cytosine in all the cytosine-amino acid dimer complexes follows a doorway mechanism, where the electron gets transferred from initial dipole-bound doorway state to the final nucleobase-bound state through the mixing of electronic and nuclear degrees of freedom. In higher amino acid concentration, amino acid-bound state acts as the doorway state, where the initial electron density is localized on the amino acid, away from the nucleobase. This leads to the physical shielding of nucleobase from the incoming extra electron. At the same time, the presence of amino acids can increase the stability of nucleobase-bound anionic state, which can suppress the dissociative electron attachment induced sugar-phosphate bond breaking.

*achintya@chem.iitb.ac.in

# 1. Introduction

Ionizing (X-rays, Gamma rays) and particulate radiations (electrons, protons, neutrons, beta, and alpha particles) have been known for their damaging effect on DNA for many years.[1] One of the major applications of these phenomena is in the field of cancer treatment, where ionizing radiations are used to kill tumor cells. A major part of the radiation damage to genetic material is caused by secondary processes.[2,3] The ionizing radiations, even with 1 MeV of deposited energy, can generate a large number (~ $5 * 10^4$) of secondary electrons[4] along with an array of other secondary products such as free radicals and cations[5] which are potentially harmful to DNA. A significant amount of work has been done in recent years to understand the role of secondary electrons in radiation damage to genetic materials.[3,6–9] They have been shown to play a major role in various DNA damage processes such as base release, lesions, strand breaks, etc.[10–24]

The electrons generated by the action of ionizing radiation, collide inelastically with the solvent molecules around DNA, and rapidly loose their kinetic energy to form low energy elctrons (LEE), pre-solvated and aqueous electrons in a step-wise manner.[9,25,26] The LEE and pre-solvated electrons have been considered as one of the major contributors in the radiation-induced DNA damage.[10] Starting from the first experiment of Sanche's group,[10] various experiments[11–15] and simulations[27–32] have been reported on electron attachment-induced radiation damage.

Numerous attempts[27–29,33] were made to explain the exact mechanism of initial electron attachment to the DNA. A majority of the existing theoretical reports suggest a resonance-based pathway.[10,34–38] Here, the incoming electron initially attaches to a π* orbital of the nucleobase, forming a shape or feshbach resonance state.[39–42] The electron from the π*-type resonance state subsequently gets transferred to the sugar-phosphate σ* (C–O) or sugar-nucleobase (C–N) σ* state, resulting into the C–O/C–N bond cleavage.[43–50]

We have recently reported a doorway mechanism of electron attachment to DNA nucleobases.[24,25,51] The initial electron attachment to nucleobases in the gas phase leads to the formation of a dipole-bound state, which acts as a doorway for electron capture. This has also been described previously by Sommerfeld in their earlier work.[52–54] The dipole-bound state subsequently gets converted into a valence-bound state. The potential energy surface of the ground and the first excited state of the DNA base anion shows an avoided crossing, and the transfer of electron takes place through a non-adiabatic process. A similar doorway mechanism has been observed for the nucleobase anion in the condensed phase, where the water-bound state acts as the doorway state.[25,55,56] The formation of stable anion through the doorway mechanism will be competitive with the resonance-based dissociative electron attachment in DNA.

The presence of other molecular components in close proximity to DNA, especially the proteins, can play a crucial role in determining the relative rate of the two pathways as well as the amount of induced damage. Numerous experimental and theoretical investigations have been reported in the literature on the role of protein on the radiation damage pathway. Solomun *et. al.*[57] have made the first attempt to study the role of a protein in LEE (3 eV) induced DNA strand break using fluorescence spectroscopy. They have reported that the protein is an effective inhibitor of radiation-induced DNA strand break. Experimental studies of arginine and glycine with a short DNA chain have also shown that a high concentration of amino acids in the samples results into protective action against LEE irradiation.[58] Additionally, gas phase studies on uracil-alanine and uracil-glycine complexes have observed a barrier-free proton transfer between anionic nucleobase and amino acids to stabilize excess negative charge.[59,60] Kohanoff and co-workers[8,61] have also revealed that the presence of glycine environment acts as a protective shield against the electron attachment to thymine.

However, they have restricted their attention only to the ground state of the anion. Experimental studies have shown that the excited states of the anion play an important role in the electron attachment process to the DNA[62]. The accurate calculation of the ground and excited state of a complete DNA strand is not computationally feasible. The model systems can be helpful in getting an understanding of the role of protein in electron attachment-induced damages. In this study, we have investigated the mechanism of electron attachment to cytosine in the micro-solvated and bulk amino acid solution.

## 2. Computational Details

The choice of cytosine as a nucleobase in the model system goes to its highest vertical electron affinity among all the nucleobases.[63,64] In the micro-solvation study, we have restricted our attention to cytosine-amino acid dimer complexes. Geometries of all the complexes were optimized using the RI-MP2/def2-TZVP level of theory. The cartesian coordinates for all the optimized dimers at neutral and anionic geometries are provided in the supporting information. The EA-EOM-DLPNO-CCSD[65]/aug-cc-pVXZ (X=D, T) level of theory has been used to calculate single-point electron affinities. Additional 5s5p4d diffuse functions[66] were added in an even-tempered way to the positive end of the dipole moment vector. The calculated electron affinity values have been extrapolated to the complete basis set (CBS) limit. The RIJK approximation in ORCA has been used for Hartree-Fock calculations, and the NORMALPNO setting has been used for the DLPNO calculations.[67–74]

To understand the behaviour of cytosine in a bulk amino acid medium, the cytosine molecule was placed in a cubic box of 40Å length and solvated with 827 glycine molecules. The number of glycine molecules to be put inside the box was calculated from the equilibrium density of glycine at room temperature. Classical molecular dynamics (MD) simulations were performed using the NAMD[75] software package to generate the distribution of glycine molecules around the cytosine molecule. A CHARMM compatible forcefield, parameterized with SwissParam[76] has been used for both cytosine and glycine. These forcefield files can be downloaded from our website.[77]

The system was minimized initially with a non-bonded interaction cut-off of 12 Å, and positional constraints were applied on cytosine. After minimization, the system was slowly heated to 300 K and then equilibrated for 10 ns at a constant temperature and pressure of 300 K and 1 atmosphere, respectively. Periodic boundary conditions and particle mesh Ewald technique (PME) have been employed in the equilibration run. Subsequently, a 30 ns production run was carried out at constant volume and constant temperature (300 K). Equilibration and production run were performed with a time integration step of 2 fs.

From the 30 ns classical MD trajectory, four snapshots were taken from the last 15 ns at an interval of 5 ns i.e., at 15, 20, 25, and 30ns. These snapshots were used for the subsequent QM/MM molecular dynamics simulations. Cytosine and the surrounding glycine molecules within 2.7 Å region of the cytosine were considered as the QM region for the QM/MM MD run and treated at the BP86/def2-SVP level of theory. The remaining glycine molecules are treated at the MM level using the same forcefield as that of the classical MD simulations. The QM/MM MD simulations were performed for 2 ps with a time step of 0.5 fs on each of the four snapshots taken from the classical MD. Each trajectory was run three times to ensure proper sampling. The QM/MM calculations were performed using a box length of 40 Å without any periodic boundary conditions and non-bonded cut-offs. The size of the 2.7 Å QM region has been chosen based on the radial distribution function plot from the classical MD simulations. More details of which have been given in the supporting information (Figure S1). All the QM/MM calculations have been performed using ORCA quantum chemistry package.[78]

## 3. Results and discussion

### 3.1 Effect of micro solvation

To unravel the role of amino acids in the process of electron attachment to cytosine, four different amino acids (arginine, lysine, alanine and glycine) are chosen to form complexes with cytosine. Glycine is selected due to its nonpolar nature and the presence of a short side chain, which will keep the cost of the calculations down. Whereas arginine, alanine, and lysine are considered as they form a large part of histone proteins bonded to DNA in the biological environment. Table 1 presents the dipole moment and vertical electron affinity (VEA) corresponding to the dipole-bound state of all four complexes of cytosine and amino acid.

The addition of an electron to the neutral cytosine leads to the formation of a dipole-bound state where the excess electron is located away from the nuclear framework (See Figure 1). The geometry distortion is minimal[51] in the dipole bound case which leads to almost identical values of VEA with Vertical Detachment Energy (VDE) and Adiabatic Electron Affinity (AEA) for the dipole bound state. All four amino acids, cytosine, and the considered cytosine-amino acid complexes show the presence of a dipole-bound state.

The dipole moment and VEA values mentioned in Table 1 vary widely. Although there is no quantitative correlation between the dipole moment and VEA for dipole-bound states, the arginine-cytosine complex with the highest dipole moment among all the complexes shows a maximum VEA value of 0.144 eV. From Figure 1, one can notice that the dipole-bound state is localized towards cytosine for all four complexes. It essentially means that the presence of amino acid does not lead to any physical shielding of cytosine from electron attachment in the micro-solvated complexes.

The valence-type anionic states of the cytosine-amino acid complexes are not stable at neutral geometry and are only bound in the anionic geometry. The VDE and AEA corresponding to the valence-bound state of all the complexes are also mentioned in Table 1. For the isolated amino acid, the excess electron is not bound even at the anionic geometry, which is reflected in negative AEA values of all the amino acids. The AEA value is negative for isolated cytosine also. However, the AEA value in the case of all the complexes is positive, depicting the formation of the stable valence-bound anionic state. The isolated cytosine has a larger (less negative) AEA value than the amino acids. Hence, one can assume that the excess electron in the valence-bound state of complexes should preferentially localize on the nucleobase. This is indeed observed in all four complexes (See Figure 1). The positive AEA values in the case of complexes can be related to the stabilizing effect of hydrogen bonding between the cytosine and the amino acids. Out of all the four complexes, the VDE is maximum in the case of cytosine-arginine complex, with a value of 1.57 eV. Higher stability of this complex can be attributed to a larger number of H-bonding interaction sites in arginine compared to the other three amino acids.

The amino acids can act as a protective agent for the nucleobase in two ways.[58,61] Firstly, the amino acid can act as a protective screen if the excess electron is initially localized over it. Else, if the electron is captured initially by the nucleobase, the amino acid can stabilize the excess negative charge on the nucleobase via various interactions; hydrogen bonding being one of the most important contributors. The first pathway of protection is not observed in the case of micro-solvated structures. However, the stability of valence-bound anionic state of cytosine is considerably increased by the presence of amino acid.

The formation of valence-bound anionic state is followed by the proton transfer from amino acid to cytosine in the cases of cytosine-alanine and cytosine-glycine complex (See Figures 2 and 3). Although there is no explicit proton transfer is observed in going from neutral to the anionic state for cytosine-arginine and cytosine-lysine complexes, but the distance between the carbonyl oxygen of cytosine and the O-H or N-H proton of the amino acid considerably decreases on going to anions. It can be seen from Figures 2 and 3 that the complex having arginine as an amino acid shows a decrease in the bond distance between the carbonyl oxygen of cytosine and acidic hydrogen of arginine from 1.69 to 1.37 Å, as it moves from neutral to anion geometry.

A similar trend is observed in the case of the cytosine-lysine dimer, where the bond distance between amino hydrogen of lysine and the carbonyl oxygen of cytosine is decreased from 2.33 to 2.24 Å on going from neutral to anionic geometry. At the same time, the increase in bond length between amino hydrogen of cytosine and amino nitrogen of cytosine from 1.93 to 2.15 Å, as shown in Figures 2 and 3, shows the very nature of cytosine as a proton acceptor rather than a donor. The explicit proton transfer and stronger hydrogen bonding interaction increase the stability of the valence-bound cytosine anion in the cytosine-amino acid complexes, resulting in larger AEA values.

The initial dipole-bound states in cytosine-amino acid dimers act as a doorway for electron capture, and the electron subsequently gets transferred to the cytosine-bound state. To understand the mechanism of electron transfer from the dipole bound to the valence bound state of cytosine with and without the presence of amino acids, we have calculated the adiabatic potential energy curve (PEC) of ground and first excited state of the anion along a linear transit from dipole-bound to valence-bound geometry. Intermediate geometries along the linear transit can be derived using the following expression:

$$R_{int} = (1 - \lambda)R_{DB} + \lambda R_{VB} \qquad (1)$$

Where $R_{DB}$ is the parameter (bond length, bond angle and dihedral angle) for the dipole-bound geometry, and $R_{VB}$ is the parameter corresponding to the geometry of the valence-bound state. $\lambda$ is the linear transition parameter, which varies from 0 to 1. The $\lambda = 0$ leads to the dipole-bound geometry, and $\lambda = 1$ leads to the valence-bound geometry. The adiabatic potential energy curve in Figure 4 shows an avoided crossing between the ground and excited states of the anion, indicating a mixing of electronic and nuclear degrees of freedom. Such mixed states cannot be treated by standard Born-Oppenheimer approximation.[79]

As a potential solution, one can switch to a diabatic representation where the electro-nuclear coupling is replaced by the electronic coupling between the two diabatic states. Two diabatic states were generated using morse potential fitting to the valence-bound and dipole-bound parts of the ground state adiabatic PEC. The PEC for all the four cytosine-amino acid complexes are presented in supporting information (See Figure S3). A two-state avoided crossing model potential[79] has been used to calculate the coupling between the two diabatic states as follows:

$$V = \begin{pmatrix} V_1 & W \\ W & V_2 \end{pmatrix} \qquad (2)$$

where the diagonal elements $V_1$ and $V_2$ are morse potentials in the coordinate $\lambda$ as per the following equation, keeping off-diagonal elements constant. The potential varies with $\lambda$ as per the equation:

$$V_i = \omega_i \left(1 - e^{-a(\lambda - \lambda_i^0)}\right)^2 + V_i^0 \qquad (3)$$

One can observe from Table 2 that the calculated coupling constant value for various complexes are quite small. Therefore, one can use Marcus Theory[80] to estimate the rate of transition of an electron from dipole-bound state to valence-bound state in the weak-coupling limit as:

$$k = \frac{2\pi}{\hbar}|W|^2 \sqrt{\frac{1}{4\pi k_B T \lambda_R}} e^{-\frac{(\lambda_R + \Delta G^0)^2}{4\lambda_R k_B T}} \qquad (4)$$

where λ, W and $\Delta G^0$ represent the reorganization energy, coupling constant, and free energy change between the two states ($E_{VB}$ - $E_{DB}$), respectively.

The rate of electron transfer from dipole bound to valence-bound species is presented in Table 2 for isolated cytosine, monohydrated cytosine, and all the cytosine-amino acid complexes. It should be noted that the rate of transfer is increased to 5-6 folds in dimers in comparison to isolated cytosine. The cytosine-arginine complex shows the highest rate of electron transfer of $7.84 \times 10^{11}$ sec$^{-1}$ between the dipole and valence-bound state. The other three complexes show one-fold lower rate than that observed in the cytosine-arginine case. The rates in all four complexes are higher than that observed in the monohydrated cytosine[25]. The rapid formation of valence-bound anionic state leads to the scavenging of low-energy electrons by the nucleobase and suppresses the electron attachment induced bond breaking[24].

### 3.2 Effect of the local environment

Glycine, being the smallest of all the considered amino acids, is chosen to look into the effect of the local environment of glycine on the electron attachment to cytosine. Various conformers of the cytosine-glycine complex were generated using CREST[81] software and subsequently optimized using the XTB method[82]. The twelve lowest energy isomers were taken and further optimized at the RI-MP2/def2-TZVP level of theory. Figure 5 shows all the twelve cytosine-glycine conformers generated using the CREST package, arranged in increasing order of their relative MP2 energy with respect to the most stable conformer a1. The twelve isomers of the cytosine-glycine complex can be categorized into four classes of position isomers based on their hydrogen bonding patterns. They are denoted by a, b, c, and d in Figure 5. Further calculations were performed for the most stable isomers in their respective category (a1, b1, c1 and d1). These are shown as encircled structures in Figure 5.

Dipole moment and VEA for dipole bound state, and VDE and AEA for valence bound state of these four conformers are mentioned in Table 3. The electron attachment to all the four conformers at neutral geometry leads to the formation of dipole bound anionic state (see Figure 6). Isomer b1 shows the highest dipole moment and, consequently, the highest VEA. The ground state at the anionic geometry is valence-bound in nature. One can see proton transfer from glycine to cytosine in isomers a1 and b1 in the anionic geometry. The AEA of b1 is also higher than other isomers. We have calculated the potential energy surface corresponding to the linear transit from dipole-bound to valence-bound geometry (see Figure S4) for all the four lowest energy conformers. The corresponding coupling constant and the rate of electron transfer is presented in Table 4.

Isomer c1 shows the highest coupling constant of 19.4 meV, with the rest of the three isomers showing almost half of the corresponding value. The rate of transfer from dipole-bound to valence-bound state is maximum for isomer b1, which is one order of magnitude higher than the rest of the three isomers and similar to that observed for cytosine in GC base-pair.[83] It shows that the electron attachment properties of cytosine in cytosine-glycine dimer depends upon the local interaction between cytosine and glycine. Therefore, the conclusions drawn about the electron attachment to cytosine in the presence of glycine from micro-solvation models may not be transferrable to the cytosine in a bulk glycine environment.

### 3.3 Effect of bulk glycine environment

Although a dimer model can be helpful in understanding the mechanism of the electron attachment process, to simulate the electron attachment process to DNA in a realistic environment, one needs to consider the effect of the bulk environment. A high amino acid concentration can increase the stability of cytosine anion with the formation of an extensive hydrogen-bond network.

To understand the effect of high amino acid concentration, a model system was prepared keeping cytosine in a 40 Å cubic box of glycine and the behaviour of the system with respect to electron attachment was observed using QM/MM simulations. The details of classical and quantum mechanical simulation have been mentioned in the computational details section. Subsequently, EA-EOM-DLPNO-CCSD based QM/MM single-point calculations were performed over snapshots taken from QM(BP86/def2-SVP)/MM trajectory to look at the time evolution of the ground and the excited states of the anion. The single-point QM/MM calculations were performed using with EA-EOM-DLPNO-CCSD/aug-cc-pVDZ method, treating cytosine and glycine molecules within the 2.7 Å of the cytosine at the QM level. Additional 5s5p4d diffused functions were placed on the nitrogen atom of cytosine. The remaining glycine molecules were treated with the same forcefield as that of classical molecular dynamics.

Figure 7 reveals the sequence of events observed in QM/MM trajectories. At the very beginning, the electron is localized on the bulk glycine molecules. As the dynamics proceeds, the extra electron density gets transferred from the bulk glycine molecules to nucleobase. Therefore, the glycine-bound state acts as a doorway for the formation of a stable valence-type nucleobase-bound state. The transition from glycine-bound state to nucleobase-bond state was observed around 100-105 fs and is followed by glycine to cytosine proton transfer. This glycine to cytosine proton transfer occurs around 150 fs. To further understand the nature of the initial glycine-bound state, we have reperformed the EA-EOM-DLPNO-CCSD calculation at the snapshot from 0.5 fs by keeping the cytosine molecule as a ghost. The corresponding natural orbitals are presented in Figure S5. One can see that the VDE of the system with and without cytosine is almost identical and leads to nearly identical natural orbitals for the anionic state. Therefore, this doorway state arises primarily due to the interaction of the electron with the glycine.

Figure 8 presents the time evolution of the VDE for the lowest anionic state of cytosine in bulk glycine. Here, one can see a sharp increase in the VDE value for the initial steps till 200 fs depicting the progressive stabilization of the anionic state. This could be attributed to the formation of a nucleobase bound state, which is being stabilized by proton transfer from one of the surrounding solvent glycine molecules. The VDE values appear to get converged around 1500 fs and show a value around ~6 eV, which is much larger than that observed in any of the micro-solvated structures. The proton transferred from the glycine is observed to be attached to two different sites of cytosine in different trajectories. One on the carbonyl oxygen and the other at the imine nitrogen of cytosine molecule. Figure 9 presents the radial distribution function plot for the (a) acidic hydrogen of glycine and the carbonyl oxygen of cytosine (b) acidic hydrogen of glycine and imine nitrogen of cytosine. The radial distribution function was calculated from the QM/MM simulation trajectory for the neutral and anionic cytosine in bulk solvated glycine molecules.

Here, one can see a sharp peak around 1 Å in case of interaction observed in case a and around 3 Å for case b, which indicates that a short-range local structure of glycine is formed around the cytosine anion. This local structure of the glycine distribution leads to the preferential stabilization of the cytosine anion through proton transfer. This kind of protection can be referred to as a chemical mechanism of protection, where the glycine, by preferential stabilization of the valence-bound cytosine anion, increases its electron scavenging effect and reduces the availability of secondary electrons for resonance-induced strand breaking. One should note that, unlike the micro-solvated cytosine-amino acid system, where the dipole bound state has electron density situated towards cytosine, the bulk solvated system shows stabilization via physical shielding also. The initial electron density is localized on glycine and away from the nucleobase. It reduces the possibility of base-centered resonance states[84–86] which leads to DNA strand breaking. A similar trend has been observed in all the considered QM/MM dynamics trajectories generated from other snapshots. The radial distribution function depicted in Figure 9 has been generated from the 25 ns snapshot trajectory.

## 4. Conclusions

In this work, we have studied the effect of amino acids on the electron attachment behaviour of cytosine. Four dimer complexes of amino acids (arginine, lysine, alanine, and glycine) were generated with cytosine nucleobase. In the micro-solvated model, the initial electron attachment leads to the formation of dipole-bound states where the electron density is away from the nucleobase. The electron subsequently gets transferred to the cytosine-centered valence bound state due to the mixing of electronic and nuclear degrees of freedom. The presence of amino acid stabilizes the cytosine-centered anion by hydrogen bonding and/or proton transfer and by enhancing its electron scavenging activity. However, the micro-solvated model does not show any sign of physical screening by amino acid molecules.

The situation changes on going to the bulk amino acid environment. Our calculations show that a similar doorway mechanism exists for cytosine in a bulk solution of the amino acid system also, where the glycine-bound states act as a doorway for the nucleobase-bound anionic state. The formation of a glycine-bound state keeps the initial electron density away from the nuclear framework and hence, reduces the possibility of formation of base-centered resonances, responsible for strand breaking. Here, the physical screening of electrons by the amino acids is also observed in addition to the chemical protection mechanism already observed in the micro-solvated model. As the dynamics proceeds, the stability of the cytosine-centered bound state anion increases drastically in a bulk glycine environment. Thus, by increasing the electron scavenging property of the nucelobase, it further reduces the availability of electrons for dissociative electron attachment. It will be interesting to see the effect of longer amino acid chains and the secondary structure of protein on larger DNA model systems. Work is in progress towards that direction.

**Supporting Information**

Cartesian coordinates of the optimized neutral and anionic geometry of Cytosine, Amino acids and all the Cytosine-Amino acid complexes; RMSD values from neutral to anion structure for various molecules considered and individual VDE values corresponding to the time evolution of lowest anionic state in QM/MM trajectory at different timesteps are provided in the supporting information.

**Acknowledgments**

The authors acknowledge the support from the IIT Bombay, IIT Bombay Seed Grant project, SERB CRG, and Matrix project, CSIR, ISRO, DST-Inspire Faculty Fellowship for financial support, IIT Bombay super computational facility, and C-DAC Supercomputing resources (PARAM Yuva-II & Param Brahma) for computational time.

*Table 1: The VEA corresponding to the dipole-bound state, VDE and AEA values corresponding to the valence-bound states of bare cytosine, isolated amino acids, and all the cytosine-amino acid complexes at EA-EOM-DLPNO-CCSD/CBS level of theory. The dipole moment values are calculated at the Hartree-Fock level.*

| COMPLEX | Dipole Moment (Debye) | VEA (eV) | VDE (eV) | AEA[a] (eV) |
|---|---|---|---|---|
| Cytosine | 7.29 | 0.060 | 0.08 | -0.25 |
| Arginine | 4.09 | 0.042 | 0.64 | -0.56 |
| Lysine | 3.71 | 0.021 | 0.44 | -0.78 |
| Alanine | 4.21 | 0.024 | 0.39 | -0.84 |
| Glycine | 4.24 | 0.031 | 0.45 | -0.99 |
| Cytosine + Arginine | 7.65 | 0.144 | 1.57 | 0.57 |
| Cytosine + lysine | 6.75 | 0.073 | 1.16 | 0.18 |
| Cytosine + Alanine | 6.33 | 0.058 | 1.47 | 0.30 |
| Cytosine + Glycine (iso **a1**) | 6.73 | 0.065 | 1.44 | 0.36 |

[a]Zero point energy correction calculated at RI-MP2/def2-SVP level of theory has been added to the AEA.

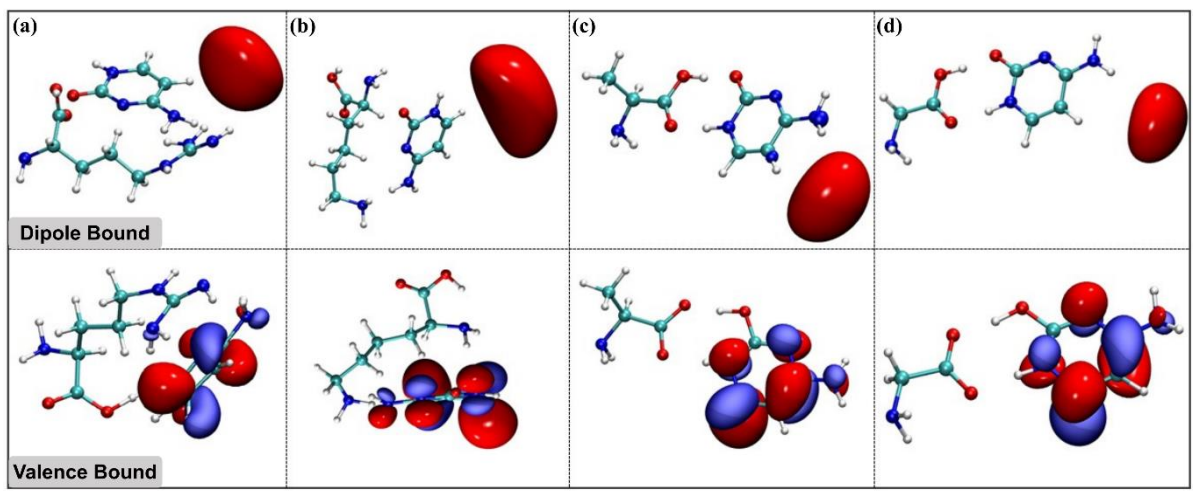

*Figure 1: Natural orbitals corresponding to dipole and valence bound state of four complexes of a nucleobase (cytosine) and amino acid (a) arginine, (b) lysine, (c) alanine and (d) glycine.*

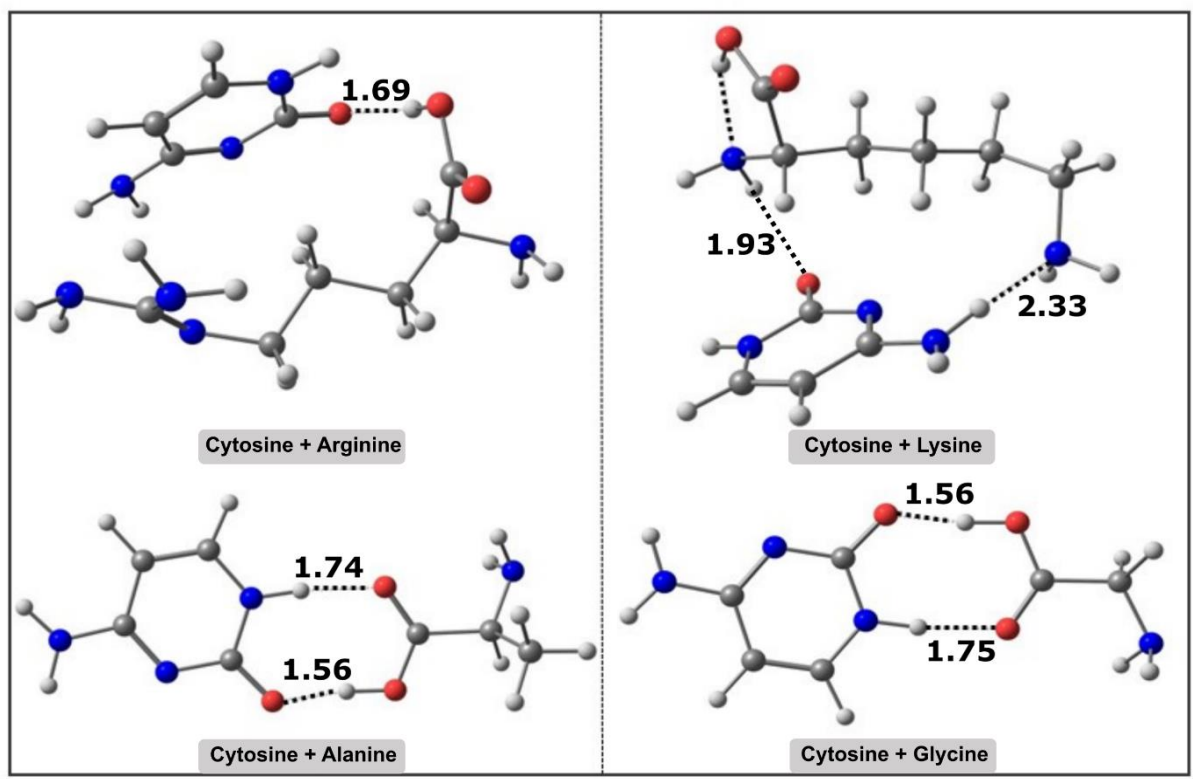

*Figure 2: Optimized structures of four cytosine + amino acid complexes at neutral geometry.*

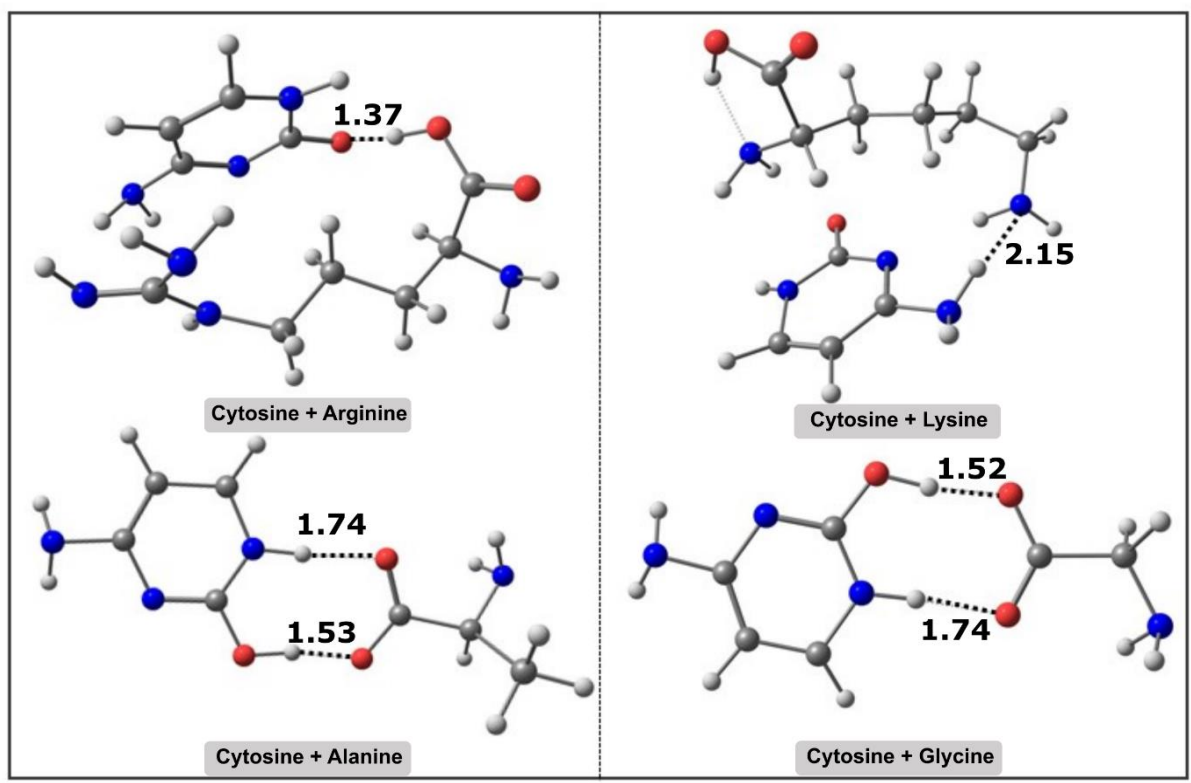

*Figure 3: Optimized structures of four cytosine + amino acid complexes at anion geometry.*

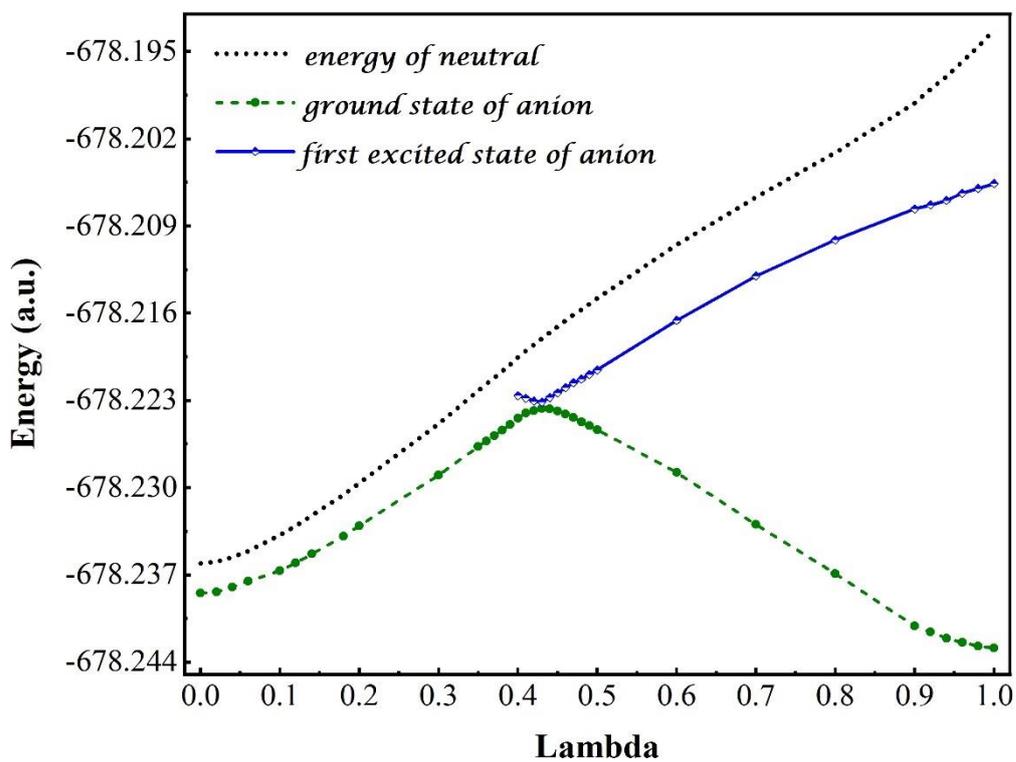

*Figure 4: Adiabatic potential energy curve of the ground and first excited state of cytosine-glycine complex (isomer a1) at EA-EOM-DLPNO-CCSD/aug-cc-pVTZ.*

*Table 2: Rate of electron transfer from dipole bound to valence bound state at room temperature along with the coupling constant (W (meV)).*

| Molecule | Coupling Constant (W) | Rate (sec$^{-1}$) |
|---|---|---|
| Cytosine + arginine | 24.14 | $7.84 * 10^{11}$ |
| Cytosine + lysine | 20.76 | $1.21 * 10^{10}$ |
| Cytosine + alanine | 10.02 | $1.30 * 10^{10}$ |
| Cytosine + Glycine (iso a1) | 8.81 | $2.69 * 10^{10}$ |
| Monohydrated Cytosine[a] | 19.83 | $1.52 * 10^{9}$ |
| Bare Cytosine[b] | 12.56 | $1.6 * 10^{5}$ |

a Taken from ref25

b Taken from ref51

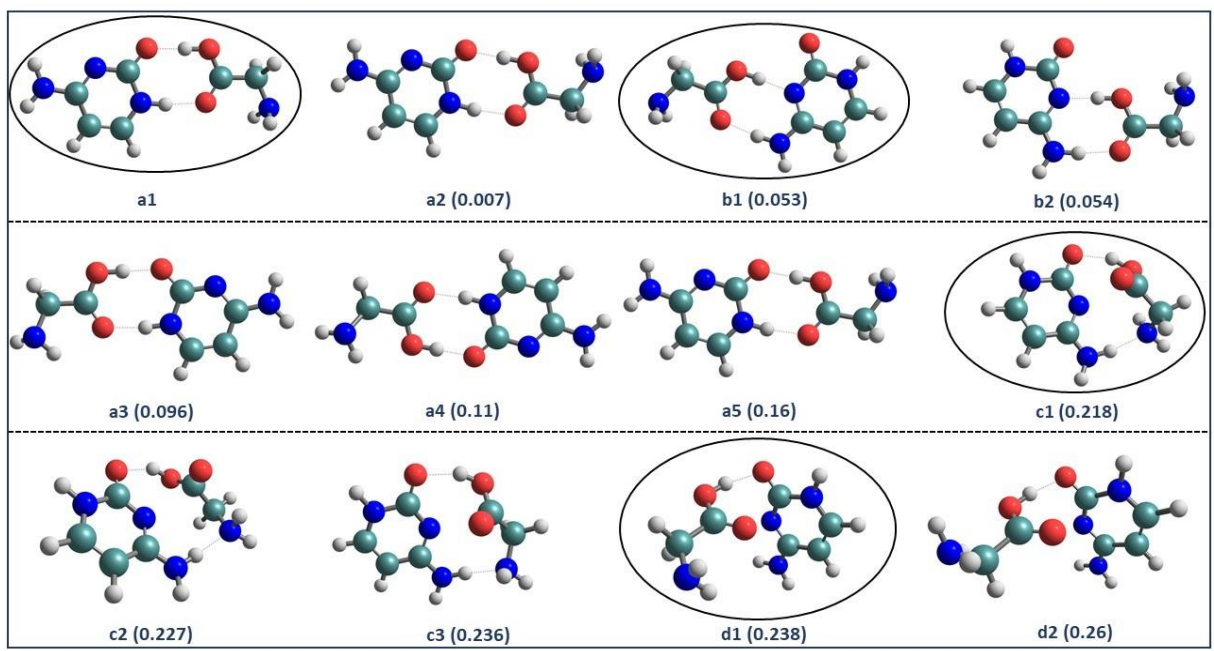

*Figure 5: The twelve lowest lying cytosine + glycine conformers. The relative energy (in kcal/mol) with respect to the lowest energy isomer (a1) at the RI-MP2 level of theory is provided. The most stable isomers in a particular class of position isomers are marked in circle.*

*Table 3: The VEA, VDE and AEA values of all the four cytosine-glycine position isomers at EA-EOM-DLPNO-CCSD/CBS level of theory. The dipole moment values are calculated at the Hartree-Fock level.*

| COMPLEX | Dipole Moment | VEA (eV) | VDE (eV) | AEA (eV) |
|---|---|---|---|---|
| Cytosine + Glycine (iso **a1**) | 6.73 | 0.065 | 1.44 | 0.36 |
| Cytosine + Glycine (iso **b1**) | 7.31 | 0.091 | 1.91 | 0.66 |
| Cytosine + Glycine (iso **c1**) | 5.35 | 0.038 | 1.39 | 0.31 |
| Cytosine + Glycine (iso **d1**) | 6.60 | 0.075 | 1.30 | 0.30 |

[a]Zero point energy correction calculated at RI-MP2/def2-SVP level of theory has been added to the AEA.

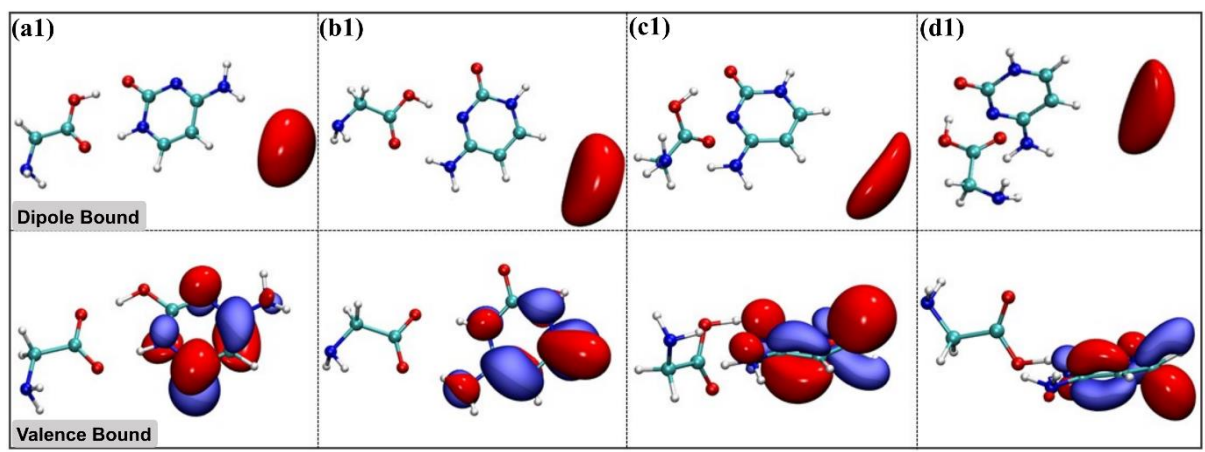

*Figure 6: Natural orbitals corresponding to dipole and valence bound state of four isomers of Glycine + Cytosine.*

*Table 4: Rate of electron transfer from dipole bound to valence bound state at room temperature along with coupling constant (W (meV)) for all the four lowest energy cytosine-glycine position isomers.*

| Molecule | Coupling Constant (W) | Rate (sec$^{-1}$) |
|---|---|---|
| Cytosine + Glycine (iso a1) | 8.81 | $2.69 * 10^{10}$ |
| Cytosine + Glycine (iso b1) | 9.65 | $1.00 * 10^{11}$ |
| Cytosine + Glycine (iso c1) | 19.40 | $1.70 * 10^{10}$ |
| Cytosine + Glycine (iso d1) | 10.55 | $2.49 * 10^{10}$ |

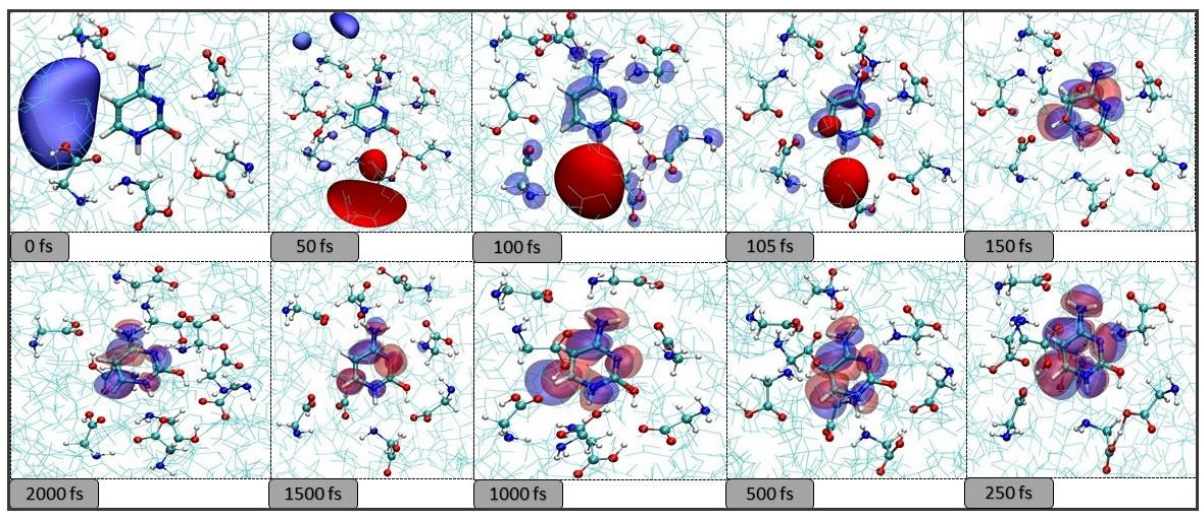

*Figure 7: Evolution of lowest anionic state of cytosine in the presence of bulk glycine environment.*

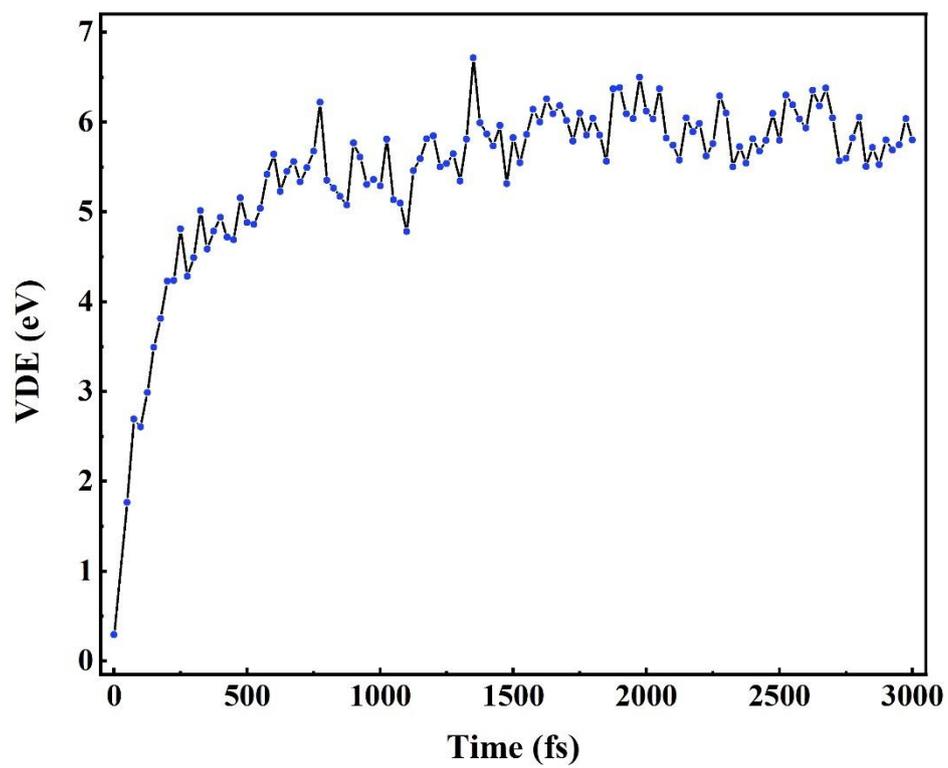

*Figure 8: Time evolution of detachment energy of lowest anionic state in QM/MM trajectory.*

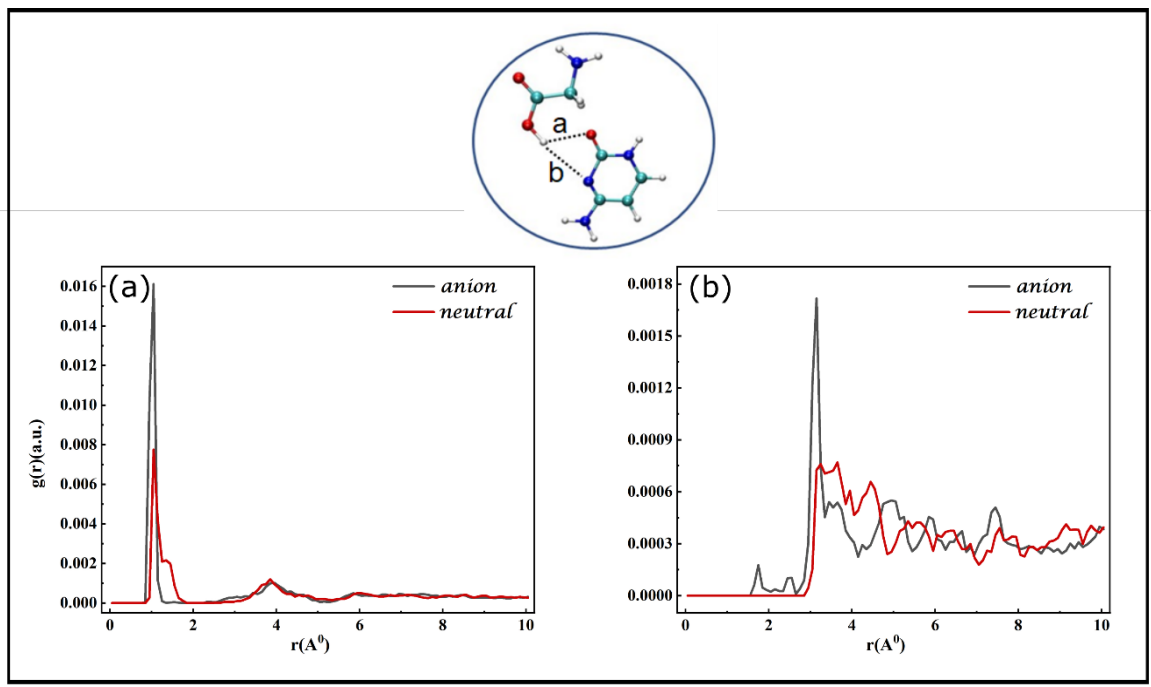

*Figure 9: Radial distribution function plot of QM/MM trajectory.*

*(a) carbonyl oxygen of cytosine and acidic hydrogen of glycine (b) imine nitrogen of cytosine and acidic hydrogen of glycine.*

# References

(1) Obodovskiy, I. M. Radiation Therapy. *Radiation* **2019**.

(2) O'Neill, P. Radiation-Induced Damage in DNA. In *Radiation Chemistry*; Jonah, C. D., Rao, B. S. M., Eds.; Studies in Physical and Theoretical Chemistry; Elsevier, 2001; Vol. 87, pp 585–622. https://doi.org/https://doi.org/10.1016/S0167-6881(01)80023-9.

(3) Alizadeh, E.; Sanche, L. Precursors of Solvated Electrons in Radiobiological Physics and Chemistry. *Chem Rev* **2012**, *112* (11), 5578–5602. https://doi.org/10.1021/cr300063r.

(4) Bichsel, H.; Pierson, D. H.; Boring, J. W.; Green, A.; Inokuti, M.; Hurst, G. *ICRU Report 31: Average Energy Required to Produce an Ion Pair*; 1979.

(5) Alizadeh, E.; Sanche, L. Precursors of Solvated Electrons in Radiobiological Physics and Chemistry. *Chem Rev* **2012**, *112* (11), 5578–5602. https://doi.org/10.1021/cr300063r.

(6) Baccarelli, I.; Bald, I.; Gianturco, F. A.; Illenberger, E.; Kopyra, J. Electron-Induced Damage of DNA and Its Components: Experiments and Theoretical Models. *Phys Rep* **2011**, *508* (1), 1–44. https://doi.org/https://doi.org/10.1016/j.physrep.2011.06.004.

(7) Alizadeh, E.; Orlando, T. M.; Sanche, L. Biomolecular Damage Induced by Ionizing Radiation: The Direct and Indirect Effects of Low-Energy Electrons on DNA. *Annu Rev Phys Chem* **2015**, *66* (1), 379–398. https://doi.org/10.1146/annurev-physchem-040513-103605.

(8) Kohanoff, J.; McAllister, M.; Tribello, G. A.; Gu, B. Interactions between Low Energy Electrons and DNA: A Perspective from First-Principles Simulations. *Journal of Physics: Condensed Matter* **2017**, *29* (38), 383001. https://doi.org/10.1088/1361-648X/aa79e3.

(9) Kumar, A.; Becker, D.; Adhikary, A.; Sevilla, M. D. Reaction of Electrons with DNA: Radiation Damage to Radiosensitization. *Int J Mol Sci* **2019**, *20* (16). https://doi.org/10.3390/ijms20163998.

(10) Boudaïffa, B.; Cloutier, P.; Hunting, D.; Huels, M. A.; Sanche, L. Resonant Formation of DNA Strand Breaks by Low-Energy (3 to 20 EV) Electrons. *Science (1979)* **2000**, *287* (5458), 1658–1660. https://doi.org/10.1126/science.287.5458.1658.

(11) Huels, M. A.; Boudaïffa, B.; Cloutier, P.; Hunting, D.; Sanche, L. Single, Double, and Multiple Double Strand Breaks Induced in DNA by 3−100 EV Electrons. *J Am Chem Soc* **2003**, *125* (15), 4467–4477. https://doi.org/10.1021/ja029527x.

(12) Abdoul-Carime, H.; Gohlke, S.; Fischbach, E.; Scheike, J.; Illenberger, E. Thymine Excision from DNA by Subexcitation Electrons. *Chem Phys Lett* **2004**, *387* (4), 267–270. https://doi.org/https://doi.org/10.1016/j.cplett.2004.02.022.

(13) Zheng, Y.; Cloutier, P.; Hunting, D. J.; Sanche, L.; Wagner, J. R. Chemical Basis of DNA Sugar–Phosphate Cleavage by Low-Energy Electrons. *J Am Chem Soc* **2005**, *127* (47), 16592–16598. https://doi.org/10.1021/ja054129q.

(14) Sanche, L. Low Energy Electron-Driven Damage in Biomolecules. *The European Physical Journal D - Atomic, Molecular, Optical and Plasma Physics* **2005**, *35* (2), 367–390. https://doi.org/10.1140/epjd/e2005-00206-6.

(15) Zheng, Y.; Cloutier, P.; Hunting, D. J.; Wagner, J. R.; Sanche, L. Phosphodiester and N-Glycosidic Bond Cleavage in DNA Induced by 4–15 EV Electrons. *J Chem Phys* **2006**, *124* (6), 64710. https://doi.org/10.1063/1.2166364.

(16) Wang, C.-R.; Nguyen, J.; Lu, Q.-B. Bond Breaks of Nucleotides by Dissociative Electron Transfer of Nonequilibrium Prehydrated Electrons: A New Molecular Mechanism for Reductive DNA Damage. *J Am Chem Soc* **2009**, *131* (32), 11320–11322. https://doi.org/10.1021/ja902675g.

(17) Li, Z.; Cloutier, P.; Sanche, L.; Wagner, J. R. Low-Energy Electron-Induced DNA Damage: Effect of Base Sequence in Oligonucleotide Trimers. *J Am Chem Soc* **2010**, *132* (15), 5422–5427. https://doi.org/10.1021/ja9099505.

(18) Kopyra, J. Low Energy Electron Attachment to the Nucleotide Deoxycytidine Monophosphate: Direct Evidence for the Molecular Mechanisms of Electron-Induced DNA Strand Breaks. *Phys. Chem. Chem. Phys.* **2012**, *14* (23), 8287–8289. https://doi.org/10.1039/C2CP40847C.

(19) Kouass Sahbani, S.; Sanche, L.; Cloutier, P.; Bass, A. D.; Hunting, D. J. Loss of Cellular Transformation Efficiency Induced by DNA Irradiation with Low-Energy (10 EV) Electrons. *J Phys Chem B* **2014**, *118* (46), 13123–13131. https://doi.org/10.1021/jp508170c.

(20) Shao, Y.; Dong, Y.; Hunting, D.; Zheng, Y.; Sanche, L. Unified Mechanism for the Generation of Isolated and Clustered DNA Damages by a Single Low Energy (5–10 EV) Electron. *The Journal of Physical Chemistry C* **2017**, *121* (4), 2466–2472. https://doi.org/10.1021/acs.jpcc.6b12110.

(21) Khorsandgolchin, G.; Sanche, L.; Cloutier, P.; Wagner, J. R. Strand Breaks Induced by Very Low Energy Electrons: Product Analysis and Mechanistic Insight into the Reaction with TpT. *J Am Chem Soc* **2019**, *141* (26), 10315–10323. https://doi.org/10.1021/jacs.9b03295.

(22) Dong, Y.; Gao, Y.; Liu, W.; Gao, T.; Zheng, Y.; Sanche, L. Clustered DNA Damage Induced by 2–20 EV Electrons and Transient Anions: General Mechanism and Correlation to Cell Death. *J Phys Chem Lett* **2019**, *10* (11), 2985–2990. https://doi.org/10.1021/acs.jpclett.9b01063.

(23) Dong, Y.; Liao, H.; Gao, Y.; Cloutier, P.; Zheng, Y.; Sanche, L. Early Events in Radiobiology: Isolated and Cluster DNA Damage Induced by Initial Cations and Nonionizing Secondary Electrons. *J Phys Chem Lett* **2021**, *12* (1), 717–723. https://doi.org/10.1021/acs.jpclett.0c03341.

(24) Narayanan S J, J.; Tripathi, D.; Dutta, A. K. Doorway Mechanism for Electron Attachment Induced DNA Strand Breaks. *J Phys Chem Lett* **2021**, *12* (42), 10380–10387. https://doi.org/10.1021/acs.jpclett.1c02735.

(25) Verma, P.; Ghosh, D.; Dutta, A. K. Electron Attachment to Cytosine: The Role of Water. *Journal of Physical Chemistry A* **2021**, *125* (22), 4683–4694. https://doi.org/10.1021/acs.jpca.0c10199.

(26) Alizadeh, E.; Sanche, L. Precursors of Solvated Electrons in Radiobiological Physics and Chemistry. *Chem Rev* **2012**, *112* (11), 5578–5602. https://doi.org/10.1021/cr300063r.

(27) Berdys, J.; Anusiewicz, I.; Skurski, P.; Simons, J. Theoretical Study of Damage to DNA by 0.2–1.5 EV Electrons Attached to Cytosine. *J Phys Chem A* **2004**, *108* (15), 2999–3005. https://doi.org/10.1021/jp035957d.

(28) Berdys, J.; Anusiewicz, I.; Skurski, P.; Simons, J. Damage to Model DNA Fragments from Very Low-Energy (<1 EV) Electrons. *J Am Chem Soc* **2004**, *126* (20), 6441–6447. https://doi.org/10.1021/ja049876m.

(29) Anusiewicz, I.; Berdys, J.; Sobczyk, M.; Skurski, P.; Simons, J. Effects of Base π-Stacking on Damage to DNA by Low-Energy Electrons. *J Phys Chem A* **2004**, *108* (51), 11381–11387. https://doi.org/10.1021/jp047389n.

(30) Simons, J. How Do Low-Energy (0.1–2 EV) Electrons Cause DNA-Strand Breaks? *Acc Chem Res* **2006**, *39* (10), 772–779. https://doi.org/10.1021/ar0680769.

(31) Gu, J.; Wang, J.; Leszczynski, J. Electron Attachment-Induced DNA Single Strand Breaks:  C3'–O3' σ-Bond Breaking of Pyrimidine Nucleotides Predominates. *J Am Chem Soc* **2006**, *128* (29), 9322–9323. https://doi.org/10.1021/ja063309c.

(32) Gu, J.; Xie, Y.; Schaefer, H. F. Near 0 EV Electrons Attach to Nucleotides. *J Am Chem Soc* **2006**, *128* (4), 1250–1252. https://doi.org/10.1021/ja055615g.

(33) Barrios, R.; Skurski, P.; Simons, J. Mechanism for Damage to DNA by Low-Energy Electrons. *J Phys Chem B* **2002**, *106* (33), 7991–7994. https://doi.org/10.1021/jp013861i.

(34) Fennimore, M. A.; Matsika, S. Core-Excited and Shape Resonances of Uracil. *Phys. Chem. Chem. Phys.* **2016**, *18* (44), 30536–30545. https://doi.org/10.1039/C6CP05342D.

(35) Burrow, P. D.; Gallup, G. A.; Scheer, A. M.; Denifl, S.; Ptasinska, S.; Märk, T.; Scheier, P. Vibrational Feshbach Resonances in Uracil and Thymine. *J Chem Phys* **2006**, *124* (12), 124310. https://doi.org/10.1063/1.2181570.

(36) Sieradzka, A.; Gorfinkiel, J. D. Theoretical Study of Resonance Formation in Microhydrated Molecules. II. Thymine-(H2O)n, n = 1,2,3,5. *J Chem Phys* **2017**, *147* (3), 34303. https://doi.org/10.1063/1.4993946.

(37) Cornetta, L. M.; Coutinho, K.; Varella, M. T. do N. Solvent Effects on the Π* Shape Resonances of Uracil. *J Chem Phys* **2020**, *152* (8), 84301. https://doi.org/10.1063/1.5139459.

(38) Fennimore, M. A.; Matsika, S. Electronic Resonances of Nucleobases Using Stabilization Methods. *J Phys Chem A* **2018**, *122* (16), 4048–4057. https://doi.org/10.1021/acs.jpca.8b01523.

(39) Dora, A.; Tennyson, J.; Bryjko, L.; van Mourik, T. R-Matrix Calculation of Low-Energy Electron Collisions with Uracil. *J Chem Phys* **2009**, *130* (16), 164307. https://doi.org/10.1063/1.3119667.

(40) Fennimore, M. A.; Matsika, S. Core-Excited and Shape Resonances of Uracil. *Phys. Chem. Chem. Phys.* **2016**, *18* (44), 30536–30545. https://doi.org/10.1039/C6CP05342D.

(41) Winstead, C.; McKoy, V. Low-Energy Electron Collisions with Gas-Phase Uracil. *J Chem Phys* **2006**, *125* (17), 174304. https://doi.org/10.1063/1.2353147.

(42) Fennimore, M. A.; Matsika, S. Electronic Resonances of Nucleobases Using Stabilization Methods. *J Phys Chem A* **2018**, *122* (16), 4048–4057. https://doi.org/10.1021/acs.jpca.8b01523.

(43) Alizadeh, E.; Orlando, T. M.; Sanche, L. Biomolecular Damage Induced by Ionizing Radiation: The Direct and Indirect Effects of Low-Energy Electrons on DNA. *Annu Rev Phys Chem* **2015**, *66* (1), 379–398. https://doi.org/10.1146/annurev-physchem-040513-103605.

(44) Kumar, A.; Sevilla, M. D. The Role of Πσ* Excited States in Electron-Induced DNA Strand Break Formation: A Time-Dependent Density Functional Theory Study. *J Am Chem Soc* **2008**, *130* (7), 2130–2131. https://doi.org/10.1021/ja077331x.

(45) Gu, J.; Leszczynski, J.; Schaefer, H. F. Interactions of Electrons with Bare and Hydrated Biomolecules: From Nucleic Acid Bases to DNA Segments. *Chem Rev* **2012**, *112* (11), 5603–5640. https://doi.org/10.1021/cr3000219.

(46) Gu, J.; Xie, Y.; Schaefer, H. F. Near 0 EV Electrons Attach to Nucleotides. *J Am Chem Soc* **2006**, *128* (4), 1250–1252. https://doi.org/10.1021/ja055615g.

(47) Simons, J. How Very Low-Energy (0.1–2 EV) Electrons Cause DNA Strand Breaks. In *Advances in Quantum Chemistry*; Sabin, J. R., Brändas, E., Eds.; Academic Press, 2007; Vol. 52, pp 171–188. https://doi.org/https://doi.org/10.1016/S0065-3276(06)52008-8.

(48) Simons, J. How Do Low-Energy (0.1–2 EV) Electrons Cause DNA-Strand Breaks? *Acc Chem Res* **2006**, *39* (10), 772–779. https://doi.org/10.1021/ar0680769.

(49) Berdys, J.; Anusiewicz, I.; Skurski, P.; Simons, J. Damage to Model DNA Fragments from Very Low-Energy (<1 EV) Electrons. *J Am Chem Soc* **2004**, *126* (20), 6441–6447. https://doi.org/10.1021/ja049876m.

(50) Martin, F.; Burrow, P. D.; Cai, Z.; Cloutier, P.; Hunting, D.; Sanche, L. DNA Strand Breaks Induced by 0--4 EV Electrons: The Role of Shape Resonances. *Phys. Rev. Lett.* **2004**, *93* (6), 68101. https://doi.org/10.1103/PhysRevLett.93.068101.

(51) Tripathi, D.; Dutta, A. K. Electron Attachment to DNA Base Pairs: An Interplay of Dipole- and Valence-Bound States. *J Phys Chem A* **2019**, *123* (46), 10131–10138. https://doi.org/10.1021/acs.jpca.9b08974.

(52) Sommerfeld, T. Intramolecular Electron Transfer from Dipole-Bound to Valence Orbitals: Uracil and 5-Chlorouracil. *J Phys Chem A* **2004**, *108* (42), 9150–9154. https://doi.org/10.1021/jp049082u.

(53) Sommerfeld, T. Coupling between Dipole-Bound and Valence States: The Nitromethane Anion. *Phys. Chem. Chem. Phys.* **2002**, *4* (12), 2511–2516. https://doi.org/10.1039/B202143A.

(54) Sommerfeld, T. Dipole-Bound States as Doorways in (Dissociative) Electron Attachment. *J Phys Conf Ser* **2005**, *4*, 245–250. https://doi.org/10.1088/1742-6596/4/1/036.

(55) Mukherjee, M.; Tripathi, D.; Dutta, A. K. Water Mediated Electron Attachment to Nucleobases: Surface-Bound vs Bulk Solvated Electrons. *J Chem Phys* **2020**, *153* (4), 44305. https://doi.org/10.1063/5.0010509.

(56) Mukherjee, M.; Tripathi, D.; Brehm, M.; Riplinger, C.; Dutta, A. K. Efficient EOM-CC-Based Protocol for the Calculation of Electron Affinity of Solvated Nucleobases: Uracil as a Case Study. *J Chem Theory Comput* **2020**. https://doi.org/10.1021/acs.jctc.0c00655.

(57) Solomun, T.; Skalický, T. The Interaction of a Protein–DNA Surface Complex with Low-Energy Electrons. *Chem Phys Lett* **2008**, *453* (1), 101–104. https://doi.org/https://doi.org/10.1016/j.cplett.2007.12.078.

(58) Ptasińska, S.; Li, Z.; Mason, N. J.; Sanche, L. Damage to Amino Acid–Nucleotide Pairs Induced by 1 EV Electrons. *Phys. Chem. Chem. Phys.* **2010**, *12* (32), 9367–9372. https://doi.org/10.1039/B926267A.

(59) Dąbkowska, I.; Rak, J.; Gutowski, M.; Nilles, J. M.; Stokes, S. T.; Bowen, K. H. Barrier-Free Intermolecular Proton Transfer Induced by Excess Electron Attachment to the Complex of Alanine with Uracil. *J Chem Phys* **2004**, *120* (13), 6064–6071. https://doi.org/10.1063/1.1666042.

(60) Gutowski, M.; Dabkowska, I.; Rak, J.; Xu, S.; Nilles, J. M.; Radisic, D.; Bowen Jr, K. H. Barrier-Free Intermolecular Proton Transfer in the Uracil-Glycine Complex Induced by Excess Electron Attachment. *The European Physical Journal D - Atomic, Molecular, Optical and Plasma Physics* **2002**, *20* (3), 431–439. https://doi.org/10.1140/epjd/e2002-00168-1.

(61) Gu, B.; Smyth, M.; Kohanoff, J. Protection of DNA against Low-Energy Electrons by Amino Acids: A First-Principles Molecular Dynamics Study. *Phys. Chem. Chem. Phys.* **2014**, *16* (44), 24350–24358. https://doi.org/10.1039/C4CP03906H.

(62) Ma, J.; Kumar, A.; Muroya, Y.; Yamashita, S.; Sakurai, T.; Denisov, S. A.; Sevilla, M. D.; Adhikary, A.; Seki, S.; Mostafavi, M. Observation of Dissociative Quasi-Free Electron Attachment to Nucleoside via Excited Anion Radical in Solution. *Nat Commun* **2019**, *10* (1), 102. https://doi.org/10.1038/s41467-018-08005-z.

(63) Schiedt, J.; Weinkauf, R.; Neumark, D. M.; Schlag, E. W. Anion Spectroscopy of Uracil, Thymine and the Amino-Oxo and Amino-Hydroxy Tautomers of Cytosine and Their Water Clusters. *Chem Phys* **1998**, *239* (1), 511–524. https://doi.org/https://doi.org/10.1016/S0301-0104(98)00361-9.

(64) Tripathi, D.; Dutta, A. K. Bound Anionic States of DNA and RNA Nucleobases: An EOM-CCSD Investigation. *Int J Quantum Chem* **2019**, *119* (9), e25875. https://doi.org/https://doi.org/10.1002/qua.25875.

(65) Dutta, A. K.; Saitow, M.; Demoulin, B.; Neese, F.; Izsák, R. A Domain-Based Local Pair Natural Orbital Implementation of the Equation of Motion Coupled Cluster Method for Electron Attached States. *J Chem Phys* **2019**, *150* (16), 164123. https://doi.org/10.1063/1.5089637.

(66) Tripathi, D.; Dutta, A. K. Bound Anionic States of DNA and RNA Nucleobases: An EOM-CCSD Investigation. *Int J Quantum Chem* **2019**, *119* (9), 25875. https://doi.org/10.1002/qua.25875.

(67) Pinski, P.; Riplinger, C.; Valeev, E. F.; Neese, F. Sparse Maps—A Systematic Infrastructure for Reduced-Scaling Electronic Structure Methods. I. An Efficient and Simple Linear Scaling Local MP2 Method That Uses an Intermediate Basis of Pair Natural Orbitals. *J Chem Phys* **2015**, *143* (3), 034108. https://doi.org/10.1063/1.4926879.

(68) Liakos, D. G.; Sparta, M.; Kesharwani, M. K.; Martin, J. M. L.; Neese, F. Exploring the Accuracy Limits of Local Pair Natural Orbital Coupled-Cluster Theory. *J Chem Theory Comput* **2015**, *11* (4), 1525–1539. https://doi.org/10.1021/ct501129s.

(69) Guo, Y.; Sivalingam, K.; Valeev, E. F.; Neese, F. SparseMaps—A Systematic Infrastructure for Reduced-Scaling Electronic Structure Methods. III. Linear-Scaling Multireference Domain-Based Pair Natural Orbital N-Electron Valence Perturbation Theory. *J Chem Phys* **2016**, *144* (9), 094111. https://doi.org/10.1063/1.4942769.

(70) Dutta, A. K.; Neese, F.; Izsák, R. Towards a Pair Natural Orbital Coupled Cluster Method for Excited States. *J Chem Phys* **2016**, *145* (3), 034102. https://doi.org/10.1063/1.4958734.

(71) Guo, Y.; Riplinger, C.; Becker, U.; Liakos, D. G.; Minenkov, Y.; Cavallo, L.; Neese, F. Communication: An Improved Linear Scaling Perturbative Triples Correction for the Domain Based Local Pair-Natural Orbital Based Singles and Doubles Coupled Cluster Method [DLPNO-CCSD(T)]. *J Chem Phys* **2018**, *148* (1), 011101. https://doi.org/10.1063/1.5011798.

(72) Dutta, A. K.; Nooijen, M.; Neese, F.; Izsák, R. Exploring the Accuracy of a Low Scaling Similarity Transformed Equation of Motion Method for Vertical Excitation Energies. *J Chem Theory Comput* **2018**, *14* (1), 72–91. https://doi.org/10.1021/acs.jctc.7b00802.

(73) Brabec, J.; Lang, J.; Saitow, M.; Pittner, J.; Neese, F.; Demel, O. Domain-Based Local Pair Natural Orbital Version of Mukherjee's State-Specific Coupled Cluster Method. *J Chem Theory Comput* **2018**, *14* (3), 1370–1382. https://doi.org/10.1021/acs.jctc.7b01184.

(74) Dutta, A. K.; Neese, F.; Izsák, R. Towards a Pair Natural Orbital Coupled Cluster Method for Excited States. *J Chem Phys* **2016**, *145* (3), 34102. https://doi.org/10.1063/1.4958734.

(75) Phillips, J. C.; Braun, R.; Wang, W.; Gumbart, J.; Tajkhorshid, E.; Villa, E.; Chipot, C.; Skeel, R. D.; Kalé, L.; Schulten, K. Scalable Molecular Dynamics with NAMD. *J Comput Chem* **2005**, *26* (16), 1781–1802. https://doi.org/doi:10.1002/jcc.20289.

(76) Zoete, V.; Cuendet, M. A.; Grosdidier, A.; Michielin, O. SwissParam: A Fast Force Field Generation Tool for Small Organic Molecules. *J Comput Chem* **2011**, *32* (11), 2359–2368. https://doi.org/https://doi.org/10.1002/jcc.21816.

(77) Madhubani Mukherjee, Santosh Ranga, Pooja Verma, A. K. D. *https://achintyachemist.wixsite.com/achintya/resources*.

(78) Neese, F. Software Update: The ORCA Program System—Version 5.0. *WIREs Computational Molecular Science n/a* (n/a), e1606. https://doi.org/https://doi.org/10.1002/wcms.1606.

(79) KÖPPEL H; DOMCKE, W. C. L. S. Multimode Molecular Dynamics beyond the Born-Oppenheimer Approximation. *Adv Chem Phys* **1984**.

(80) Marcus, R. A. On the Theory of Oxidation-Reduction Reactions Involving Electron Transfer. I. *J Chem Phys* **1956**, *24* (5), 966–978. https://doi.org/10.1063/1.1742723.

(81) Pracht, P.; Bohle, F.; Grimme, S. Automated Exploration of the Low-Energy Chemical Space with Fast Quantum Chemical Methods. *Phys. Chem. Chem. Phys.* **2020**, *22* (14), 7169–7192. https://doi.org/10.1039/C9CP06869D.

(82) Bannwarth, C.; Ehlert, S.; Grimme, S. GFN2-XTB—An Accurate and Broadly Parametrized Self-Consistent Tight-Binding Quantum Chemical Method with Multipole Electrostatics and Density-Dependent Dispersion Contributions. *J Chem Theory Comput* **2019**, *15* (3), 1652–1671. https://doi.org/10.1021/acs.jctc.8b01176.

(83) Ranga, S.; Mukherjee, M.; Dutta, A. K. Interactions of Solvated Electrons with Nucleobases: The Effect of Base Pairing. *ChemPhysChem* **2020**, *21* (10), 1019–1027. https://doi.org/10.1002/cphc.202000133.

(84) Neustetter, M.; Aysina, J.; da Silva, F. F.; Denifl, S. The Effect of Solvation on Electron Attachment to Pure and Hydrated Pyrimidine Clusters. *Angew Chem Int Ed Engl* **2015**, *54* (31), 9124–9126. https://doi.org/10.1002/anie.201503733.

(85) Kočišek, J.; Pysanenko, A.; Fárník, M.; Fedor, J. Microhydration Prevents Fragmentation of Uracil and Thymine by Low-Energy Electrons. *J Phys Chem Lett* **2016**, *7* (17), 3401–3405. https://doi.org/10.1021/acs.jpclett.6b01601.

(86) Poštulka, J.; Slavíček, P.; Fedor, J.; Fárník, M.; Kočišek, J. Energy Transfer in Microhydrated Uracil, 5-Fluorouracil, and 5-Bromouracil. *J Phys Chem B* **2017**, *121* (38), 8965–8974. https://doi.org/10.1021/acs.jpcb.7b07390.